\documentclass[%
superscriptaddress,
longbibliography,
preprint,
amsmath,amssymb,
aps,
floatfix
]{revtex4-1}
\usepackage{graphicx}
\usepackage{dcolumn}
\usepackage{bm}
\usepackage{hyperref}
\usepackage{xcolor}
\usepackage{ulem}
\begin{document}


\title{Creation and Characterization of Matter-Wave Breathers}

\author{D. Luo}
\affiliation{Department of Physics and Astronomy, Rice University, Houston, Texas 77005, USA}
\author{Y. Jin}
\affiliation{Department of Physics and Astronomy, Rice University, Houston, Texas 77005, USA}
\author{J. H. V. Nguyen}
\affiliation{Department of Physics and Astronomy, Rice University, Houston, Texas 77005, USA}
\author{B. A. Malomed}
\affiliation{Department of Physical Electronics, School of Electrical Engineering, Faculty of Engineering, and Center for Light-Matter Interaction, Tel Aviv University, 6997801 Tel Aviv, Israel}
\affiliation{Instituto de Alta Investigaci\'{o}n, Universidad de Tarapac\'{a}, Casilla 7D, Arica, Chile}
\author{O. V. Marchukov}
\affiliation{Department of Physical Electronics, School of Electrical Engineering, Faculty of Engineering, and Center for Light-Matter Interaction, Tel Aviv University, 6997801 Tel Aviv, Israel}
\affiliation{Institute for Applied Physics, Technical University of Darmstadt, 64289 Darmstadt, Germany}
\author{V. A. Yurovsky}
\affiliation{School of Chemistry, Tel Aviv University, 6997801 Tel Aviv, Israel}
\author{V. Dunjko}
\affiliation{Department of Physics, University of Massachusetts Boston, Boston, Massachusetts 02125, USA}
\author{M. Olshanii}
\affiliation{Department of Physics, University of Massachusetts Boston, Boston, Massachusetts 02125, USA}
\author{R. G. Hulet}
\affiliation{Department of Physics and Astronomy, Rice University, Houston, Texas 77005, USA}
\email{randy@rice.edu}

\date{Edited Sep 2, 2020}

\begin{abstract}
We report the creation of quasi-1D excited matter-wave solitons,
``breathers", by quenching the strength of the interactions in a
Bose-Einstein condensate with attractive interactions. We characterize the resulting breathing dynamics and quantify the effects of the aspect ratio of the confining potential, the strength of the quench, and the proximity of the 1D-3D crossover for the 2-soliton breather. We furthermore demonstrate the complex dynamics of a 3-soliton breather created by a stronger interaction quench. Our experimental results, which compare well with numerical simulations, provide a pathway for utilizing matter-wave breathers to explore quantum effects in large many-body systems.
\end{abstract}

\maketitle
\begin{figure}[htbp]
	\centering
		\includegraphics[width=1\textwidth]{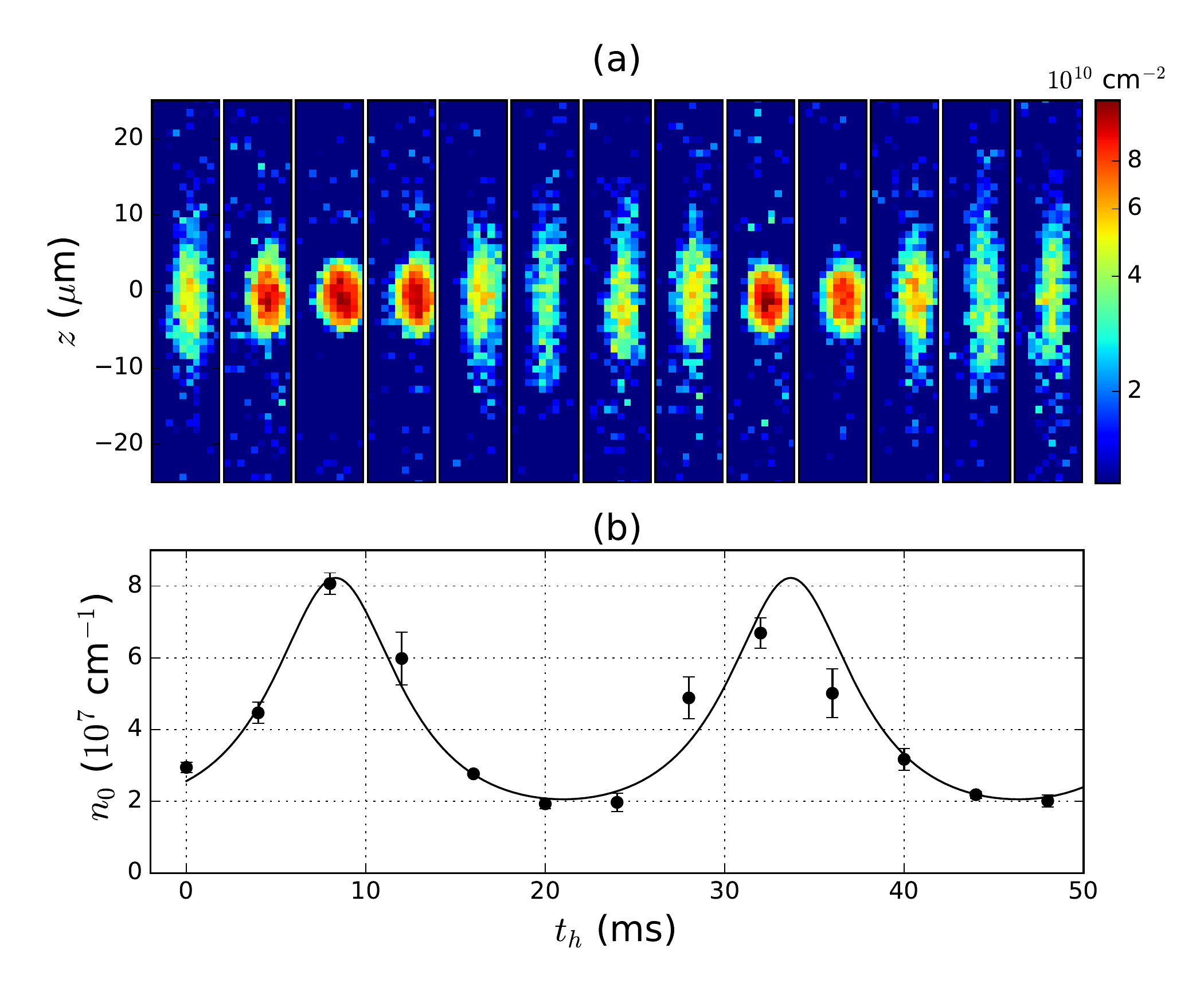}
	\caption{(a) Experimental images of a 2-soliton breather. The values of the parameters are $a_{i}=-0.15(2)\text{ }a_{0}$, $a_{f}=-0.54(3)\text{ }a_{0}$, $N=5.4(4)\times10^{4}$, $N_{c}=5.2(3)\times10^{4}$, $\omega_{r}=(2\pi)297(1)$ Hz and $\omega_{z}=(2\pi)1.12(2)$ Hz, so that $N/N_{c}=1.0(1)$, $\lambda=265(5)$, and $A^{2}=3.6(6)$. Uncertainties are discussed in Ref. \cite{Supp}. Each image is a separate realization of the experiment, and the center of the image is adjusted to remove shot-to-shot variation in the center-of-mass. The color scale represents the column density in this image, as well as in Figs. \ref{fig:Data}(c) and \ref{fig:3SolitonCD}(a). (b) Each data point is the result of fitting the axial density $n(z)$ to find its central density $n_{0}$ for each of 5 images, and averaging the result. The solid line is a fit to Eq. (\ref{eq:2solitonDensity}), with fitting parameters $\omega_{B} = (2\pi)39.4(6)$ Hz, and $\phi=(2\pi)0.17(1)$. Error bars in $n_0$ are the standard error of the mean. The uncertainty in $\omega_{B}$ is the fitting uncertainty.}
	\label{fig:2SolitonCD}
\end{figure}
\begin{figure}[htbp]
	\centering
		\includegraphics[width=1\textwidth]{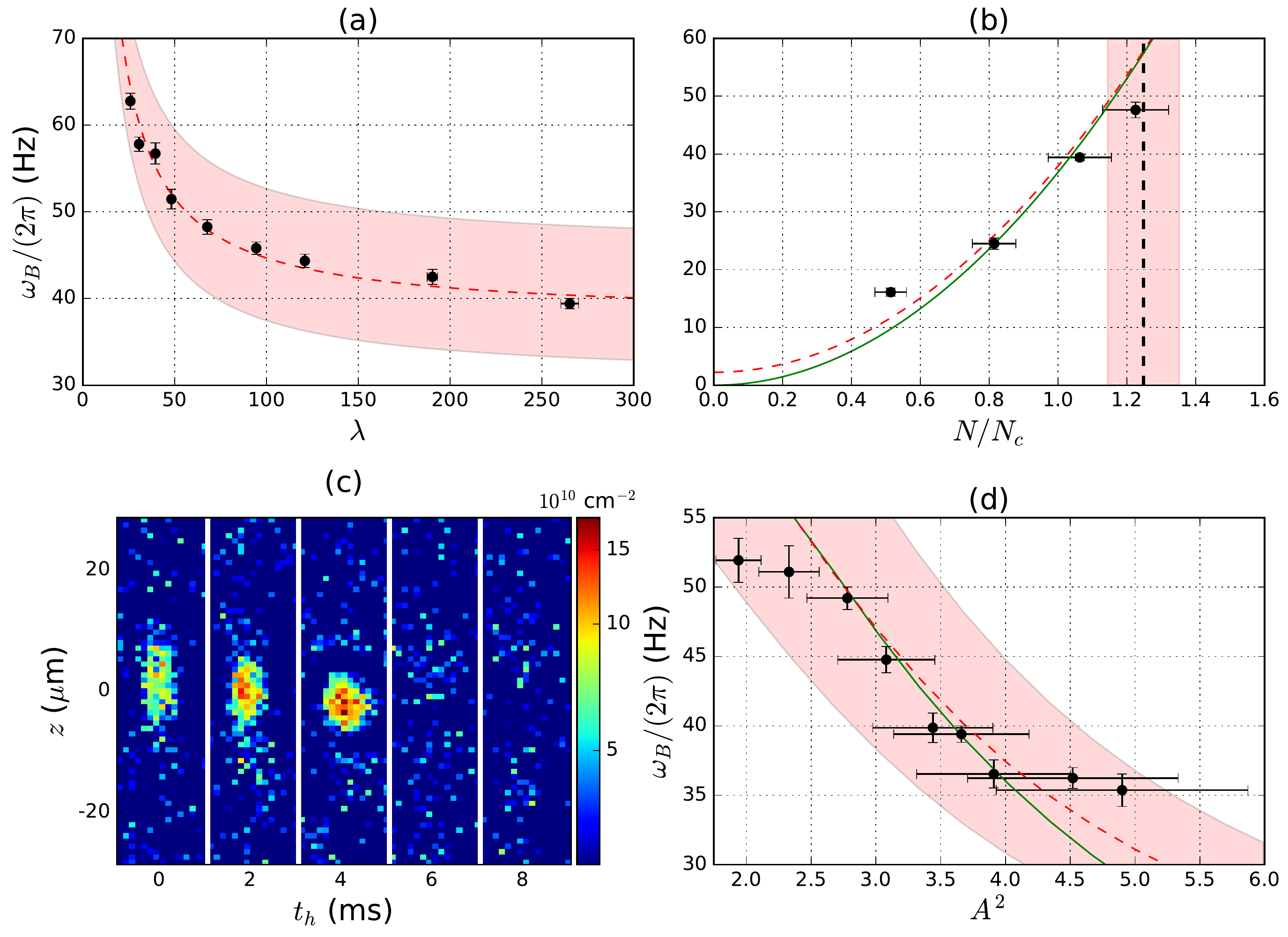}
	\caption{2-soliton breather frequency dependence on parameters. The parameters are as shown in Fig. \ref{fig:2SolitonCD} caption, unless specified otherwise. The red dashed lines in (a), (b) and (d) show the solutions of the 1D-GPE simulation, and the red shaded areas show the uncertainty range in $\omega_{B}$ due to the uncertainty in the measured $N/N_{c}$. The uncertainties for (a), (b) and (d) are discussed in Ref. \cite{Supp}. (a) $\omega_{B}$ vs $\lambda$. Here, $\omega_{r}$ is fixed while $\omega_{z}$ is varied. The location of the Feshbach resonance zero-crossing field was varied to within its uncertainty (0.2 G) to obtain the best fit GPE solutions to the data \cite{Supp}. (b) $\omega_{B}$ vs $N/N_{c}$. The solid green line is the solution to the 1D-NLSE (Eq. (\ref{eq:BreatherFreq2})). The vertical dashed line indicates the value of $N/N_{c}$ above which, collapse is observed. (c) Images showing collapse for $t_{h}$ between 4 and 6 ms after the quench and for $N/N_{c}=1.2(1)$. This sequence of images is taken from a single experimental realization. (d) $\omega_{B}$ vs $A^{2}$. Here, $a_{f}$ is fixed while $a_{i}$ is varied. The solid green line is the solution of the 1D-NLSE (Eq. (\ref{eq:Breather})).}
	\label{fig:Data}
\end{figure}
\begin{figure}[htbp]
	\centering
		\includegraphics[width=1\textwidth]{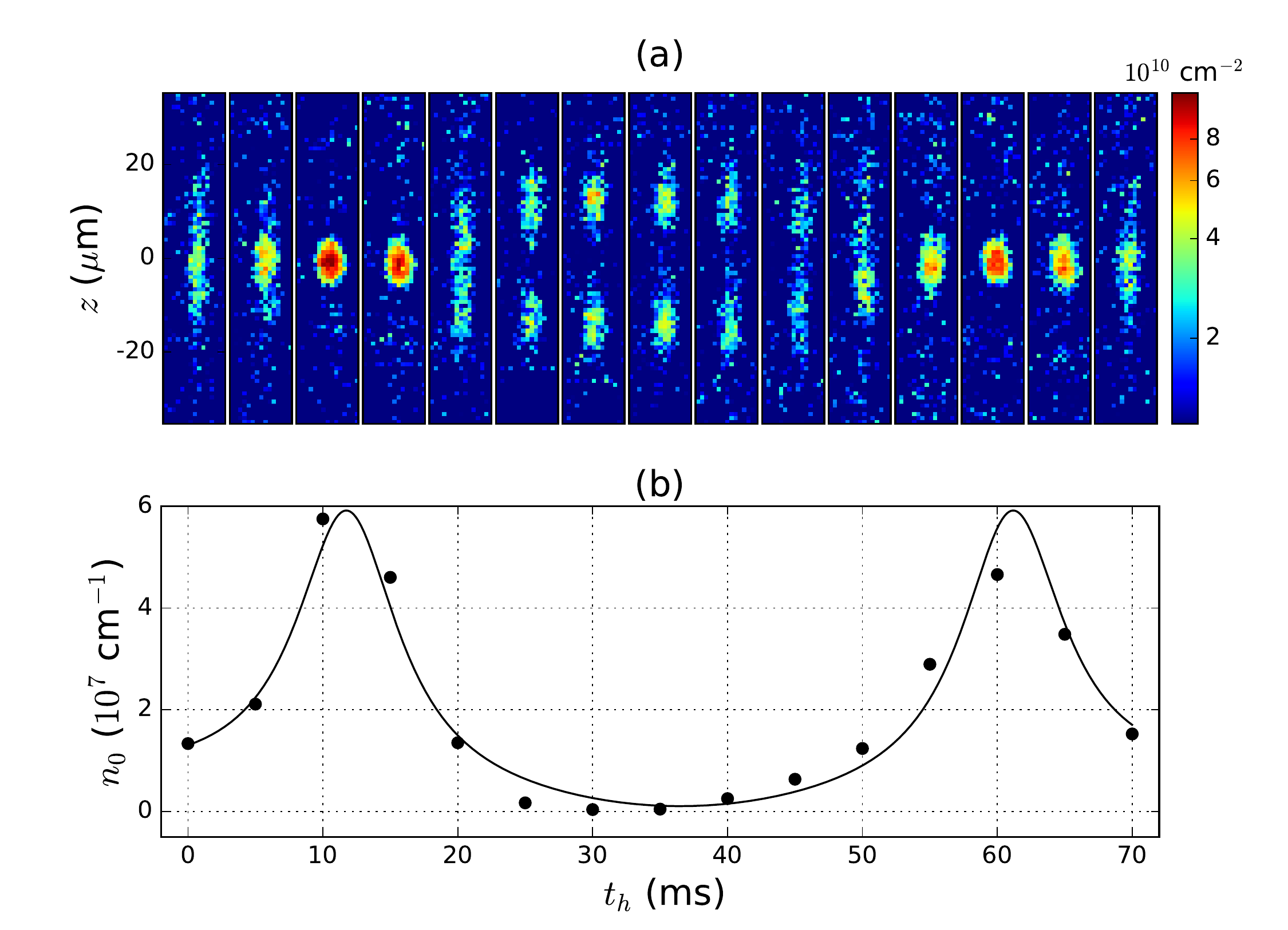}
	\caption{(a) Experimental images of a 3-soliton breather produced by $A^2=7(2)$. A series of phase-contrast images were taken at 5 ms intervals after the quench in a single realization of the experiment. The center of each image is adjusted to remove the center-of-mass variation between the images. Parameters for this data are: $\lambda = 265(5)$, $a_{i} = -0.08(2)\text{ }a_{0}, \text{ and } a_{f}=-0.57(3)\text{ }a_{0}$, and for the initial image ($t_{h}=0$), $N/N_c=1.0(1)$. Uncertainties are discussed in Ref. \cite{Supp}. In each subsequent image $N$ is reduced by  $3\%$ due to spontaneous emission by the probe. (b) The closed circles are $n_{0}$ extracted from the column density images shown in (a).  The solid line is a fit of the data to Eq. (\ref{eq:3SolitonDensity}), giving $\omega_B=(2\pi)10.6(2)$ Hz and $\phi=(2\pi)0.11(1)$.}
	\label{fig:3SolitonCD}
\end{figure}

\indent The nonlinear Schr\"{o}dinger equation (NLSE) applies to a wide variety of physical systems, such as small amplitude waves in deep water, light waves propagating in optical fiber, Langmuir waves in plasmas, and matter-waves \cite{Zakharov1980, NLSE}. A solution to the NLSE in one-dimension (1D) for a self-focusing nonlinearity is a bright soliton, a localized wave-packet that maintains its shape and amplitude while propagating. While the soliton is the ground state, the NLSE also supports excited state solutions that contain an integer number $N_{s}$ of constituent solitons. These solutions are generally supplemented by radiation that reduces the wave amplitude. In the general case, each constituent soliton is spatially separated from the others and they propagate with different velocities. A breather is a special class of an $N_{s}$-soliton where the fundamental solitons are overlapped, with zero relative-velocity, and without radiation. Unlike the case of the sine-Gordon equation, the constituent solitons of a NLSE breather are not bound to each other. Absent of any binding energy, the relative motion is in a state of neutral equilibrium \cite{Zakharov1972, Satsuma1974}. The density profile of a breather oscillates quasi-periodically with frequencies determined by the differences in the chemical potentials of the constituent solitons. The interference between the constituent solitons leads to complex spatial patterns, giving the appearance of breathing.\\
\indent Breathers were first observed in optical fiber \cite{Mollenauer1980, Stolen1983}, where optical pulses with discrete intensity levels were found to have a quasi-periodically-varying pulse-shape matching that of the $N_s=2,3$, and 4 breathers. An $N_{s}$-soliton breather can be formed from a fundamental soliton by quenching the strength of the nonlinearity by a factor of $N_{s}^2$ \cite{Satsuma1974, Carr2002}, thus creating an odd-norm-ratio breather \cite{Dunjko2014} whose fundamental solitons that form the breather have an amplitude ratio of $1:3:…:2N_{s}-1$. If the quench factor deviates from $N_{s}^{2}$, the breather becomes the closest $N_{s}$-soliton breather with a different mass ratio after shedding radiation to properly reduce the amplitude \cite{Satsuma1974}.\\
\indent In the matter-wave context, bright solitons can be formed in a Bose-Einstein condensate (BEC) confined to a quasi-1D trap by tuning the s-wave scattering length $a_{s}<0$, corresponding to an attractive nonlinearity. Matter-wave solitons, and their properties, have been the subject of intense investigation in recent years. These properties include the formation of solitons and soliton trains \cite{Strecker2002, Khaykovich2002, Eiermann2004, Cornish2006, Medley2014, Lepoutre2016, Nguyen2017, Everitt2017, MeZnarsic2019}, the collision of two solitons \cite{Nguyen2014}, interactions of solitons with potential barriers \cite{Marchant2013, Marchant2016, Wales2020} and soliton interferometry \cite{McDonald2014, Sakaguchi2016}. Solitons formed by a BEC of magnon quasi-particles in $^{3}$He have also been recently observed \cite{Autti2018a}. Recently, a 2-soliton breather was created by quenching $a_{s}$ by a factor close to 4, in combination with a rapid relaxation of the axial confinement \cite{DiCarli2019}. The soliton dynamics of these experiments are well-reproduced by the mean-field Gross-Pitaevskii equation (GPE), which is a NLSE that includes the confining potential of a trap.\\
\indent Even though the solitons in a breather spatially overlap, their binding energies are zero, leaving the relative motion of the constituent solitons sensitive to perturbations. At the same time, integrability of the NLSE protects the solitons from exchanging matter with each other or losing it to radiation. Within the framework of mean field theory, dissociation
of the breather into constituent solitons may occur due to narrow potential barriers \cite{Dunjko2014, Marchukov2019a, Grimshaw2020}. Perhaps most interestingly, beyond mean-field effects, due to quatum interference, may result in splitting \cite{Streltsov2008, Weiss2009, Weiss2012, Cosme2016}, dissociation \cite{Yurovsky2017, Marchukov2019}, relaxation \cite{Opanchuk2017, Ng2019}, or the complete lack of breathing following the quench \cite{Weiss2016}. In prior theoretical work, we evaluated the affect of quantum fluctuations on the relative velocity of the two components of a two-soliton breather using both the exact Bethe-ansatz method, appropriate for small number of atoms $N$ \cite{Yurovsky2017}, and, in the limit of large $N$, the Bogoliubov approach \cite{Marchukov2019}.  We found that quantum fluctuations can produce the macro-effect of breather dissociation over a large range of $N$, thus providing the motivation of the present study to create and characterize matter-wave breathers.\\
\indent In this work, we report the creation and characterization of a 2-soliton breather in a BEC of $^7$Li atoms, and for the first time, the experimental creation of a 3-soliton breather in a BEC. We systematically study the breathing frequency as a function of deviations from a truly 1D-system, the strength of the nonlinearity, and the quench ratio, and compare with 1D GPE simulations. We observe the characteristic dynamics of the 3-soliton breather, including density splitting and recombination, using minimally destructive sequential imaging.\\
\indent Our method for preparing an ultracold $^7$Li gas has been described previously \cite{Dries2010, Hulet2020}. The atoms are optically pumped into the $|f=1,m_{F}=1\rangle$ state, where the $s$-wave scattering length $a$ can be controlled by a broad Feshbach resonance with a zero-crossing near 544 G \cite{Pollack2009}. We describe our method for calibrating $a(B)$ in \cite{Supp}. The atoms are confined in a cylindrically-symmetric, cigar-shaped potential formed by a single-beam optical dipole trap with a $1/e^{2}$ Gaussian radius of 44 $\mu$m. In combination with axial magnetic curvature, the overall harmonic frequency along the axial ($z$) direction, $\omega_{z}$, is tunable between $(2\pi)1.12$ Hz and $(2\pi)11.50$ Hz. The radial trap frequency is $\omega_{r}=(2\pi)297$ Hz, corresponding to an aspect ratio, $\lambda=\omega_{r}/\omega_{z}$, that is between 26 and 265. We first create a BEC by direct evaporative cooling in the optical dipole trap with $\omega_{z}=(2\pi)11.50$ Hz and with $a$ tuned to 140 $a_{0}$, where $a_{0}$ is the Bohr radius. Following evaporation, we ramp $a$ from 140 $a_{0}$ to 0.1 $a_{0}$ in 1 s. During this stage, $\omega_{z}$ is kept large in order to limit the axial extent of the repulsive BEC, thus ensuring that only a single soliton is formed when the interaction is changed from repulsive to attractive. Next, $a$ is ramped from 0.1 $a_{0}$ to $a_{i}<0$ in 1 s, while simultaneously reducing $\omega_{z}$. This creates a single soliton with approximately $N=5\times10^4$ atoms, with minimal excitations. The scattering length is then quenched from $a_{i}$ to $a_{f}=A^{2}a_{i}$ in 1 ms, where $|a_{f}|>|a_{i}|$, and $A^{2}$ is the quench ratio. We use polarization phase-contrast imaging (PPCI) \cite{Bradley1997, Hulet2020} to take \textit{in-situ} images of the column density after a variable hold time $t_{h}$ following the quench.\\
\indent Figure \ref{fig:2SolitonCD} shows the breathing dynamics of a 2-soliton breather. After the quench, the wavefunction contracts towards the center and forms a large density peak at the half period, followed by expansion back to the inital profile, thus completing a full breathing period, as shown in Fig. \ref{fig:2SolitonCD}(a). The axial density $n(z)$ is obtained by integrating the column density along the remaining radial coordinate perpendicular to the imaging axis. The central density $n_{0}$ of the breather is measured by fitting the axial density to a Gaussian function $n(z)=n_{0}\exp{(-(z/l_{z})^2)}$, where $n_{0}$ and the Gaussian radius $l_{z}$ are the fitting parameters. Although $n(z)$ is not strictly a Gaussian, the $n_{0}$ found in this way is a good approximation of its true value.\\
\indent To determine the frequency of an $N_{s}$-soliton breather, the central density $n_{0}$ is measured as a function of $t_{h}$, and is fit to the corresponding analytical solution of the NLSE for 2-soliton breathers , which for $A^2=4$ is \cite{Satsuma1974}
\begin{equation}\label{eq:2solitonDensity}
n_{0}(t_{h})=\frac{\alpha}{5+3\cos{(\omega_{B}t_{h}+\phi)}},
\end{equation} 
where the breather frequency $\omega_{B}$, phase $\phi$, and overall amplitude $\alpha$ are fitted parameters. The solid line in Fig. \ref{fig:2SolitonCD}(b) shows Eq. (\ref{eq:2solitonDensity}) using the extracted parameters.\\
\indent The breather, as described by the NLSE, is a purely 1D object, while the experiment is in quasi-1D due to the fact that the ratio of the chemical potential to the radial trap frequency is non-zero, and as a result, the transverse wavefunction profile cannot be factored out. The validity of the exact NLSE breather solution also requires the absence of any axial trapping. Both the proximity to 3D and the weak axial confinement break integrability. As a consequence of being in quasi-1D, a BEC with attractive interactions is unstable to collapse once the atom number exceeds a critical value $N_{c}$. For an elongated cigar-shaped harmonic confinement, $N_{c}=0.67a_{r}/|a_{f}|$, where $a_{r}=\sqrt{\hbar/m\omega_r}=2.2$ $\mu$m is the radial harmonic oscillator length \cite{Gammal2001}. The collapse threshold for the breather is predicted to be different from that of the ground state soliton \cite{Golde2018}. We explore the 3D and axial confinement effects by measuring the dependence of $\omega_{B}$ on the trap aspect ratio $\lambda$ and, separately, on $N/N_{c}$.\\
\indent The measured $\omega_{B}$ as a function of $\lambda$ is plotted in Fig. \ref{fig:Data}(a). For this data, $N/N_{c}=1.0$, $a_{i}=-0.15\text{ }a_{0}$, and $a_{f}=-0.54\text{ }a_{0}$, giving $A^{2}=3.6$. We find that $\omega_{B}$ monotonically decreases as $\lambda$ increases from 26 to 265. We compare the measured results with the 1D Gross-Pitaevskii Equation (GPE),
\begin{equation}\label{eq:1dgpe}
i\hbar\partial_{t}\psi=-\frac{\hbar^{2}}{2m}\partial_{z}^{2}\psi+\frac{1}{2}m\omega_{z}^2 z^{2}\psi+g_{1D}N|\psi|^2\psi,
\end{equation}
where $g_{1D}=2\hbar\omega_{r}a$ is the nonlinear coupling constant \cite{Yurovsky2008b}. The ground state at $a=a_{i}$ is used as the initial wavefunction, and Eq. (\ref{eq:1dgpe}) is then numerically integrated with $a=a_{f}$ up to a few breathing periods. The resulting $\omega_{B}$, using the measured parameters, is shown by the dashed red line in Fig. \ref{fig:Data}(a). The shaded region in Fig. \ref{fig:Data}(a) represents the range of solutions of the 1D GPE that includes the measured uncertainty in $N/N_{c}$ \cite{Supp}. The measured frequency is consistent with the simulation, to within the measurement uncertainties. We also calculated $\omega_{B}$ using the 3D GPE for several values of the parameters and found excellent agreement with the 1D GPE for $N/N_{c}\lesssim0.7$. The 3D and 1D GPE differ at larger $N/N_{c}$ due to the proximity to the collapse threshold, which signals the breakdown of one-dimensionality, and eventually, of the GPE itself. Further consideration of the limits of the applicability of the 1D and 3D mean-field approximations is warranted, particularly in the case of excited states \cite{Yurovsky2002, Rogel-Salazar2002}, such as breathers.\\
\indent As mentioned above, the breather strictly exists only in 1D on a flat background, thus requiring $\omega_{B}/\omega_{z}\gg 1$. The experiment demonstrates that for $\lambda=265$, $\omega_z$ is significantly less than $\omega_B$, ensuring that the breather dynamics is indeed dominated by the nonlinear interactions, rather than the trap.\\
\indent Figure \ref{fig:Data}(b) shows the measurement of $\omega_{B}$ vs. $N/N_{c}$, for $\lambda=265$, and $A^{2}=3.6$, corresponding to the conditions to excite a 2-soliton breather. The analytic result given by the 1D-NLSE for $A^2=4$ \cite{Carr2002},
\begin{equation}\label{eq:BreatherFreq2}
\omega_{B, \text{1D}}=\frac{N^2 a_{f}^{2}}{4a_{r}^{2}}\omega_r=0.11(N/N_{c})^{2}\omega_r,
\end{equation}
is shown by the solid green curve in Fig. \ref{fig:Data}(b). The results of the 1D GPE simulation is again shown by the dashed red curve. The experimental data follows the quadratic trend given by Eq. (\ref{eq:BreatherFreq2}).\\
\indent For $N/N_{c}\geq1.2(1)$, we observe collapse of the 2-soliton breather for $t_{h}\gtrsim 4$ ms following the quench, at the time when the density grows rapidly. An example is shown in Fig. \ref{fig:Data}(c). The collapse threshold for the fundamental soliton occurs at $N/N_{c}=1.0$, which has been observed in the in-phase collisions of two fundamental solitons \cite{Nguyen2014}. A numerical simulation based on the 3D GPE \cite{Golde2018} provides an estimate of the collapse threshold for the 2-soliton breather, which is found to be $N/N_{c}=1.1$, for the experimental parameters of Fig. \ref{fig:Data}(b). Additionally, a factorization ansatz in the mean-field limit \cite{Salasnich2006} provides an analytical estimate for the collapse location to be $N/N_{c}>N_{s}^{2}/\sqrt{2N_{s}^2-1}$, which gives 1.5 for $N_{s}=2$ \cite{Supp}.\\
\indent The NLSE can predict the number of atoms in each of the two fundamental solitons when $1.5<A<2.5$. They are found to be $N_{1}=(2A-1)N/A^2$ and $N_{2}=(2A-3)N/A^2$. When $A\neq2$, the number of atoms in the two solitons, $N_{1}+N_{2}$, is less than the total number of atoms $N$, with the remaining atoms radiated away \cite{Satsuma1974}. In principle, a measurement of $N$ vs. $A^{2}$ could reveal the efficiency of the quench, but the radiated loss fraction is predicted to be less than $N/10$, and was not resolved in our experiment.\\
\indent A change in $A^{2}$ modifies the chemical potentials of the constituent solitons and, therefore, the breather frequency. The measured $\omega_{B}$ vs. the quench ratio $A^{2}$ is shown in Fig. \ref{fig:Data}(d), where the dashed red line and shaded region again correspond to the 1D GPE simulation, including uncertainties in $N/N_{c}$. The dependence of $\omega_{B}$ on $A$ for the 2-soliton breather with no axial potential can be evaluated as the soliton chemical potential difference:
\begin{equation}\label{eq:Breather}
\omega_{B, \text{1D}}(A)=\frac{16(A-1)}{A^4}\omega_{B, \text{1D}}(A=2),
\end{equation} 
which is shown by the solid green curve in Fig. \ref{fig:Data}(d).\\
\indent We also excited a 3-soliton breather by quenching by a factor of $A^2 = 7.1$.  The results are given in Fig. \ref{fig:3SolitonCD}(a), where a series of sequential images using PPCI are displayed for a single realization of the experiment. The $N_s = 3$ breather displays more complex dynamics than does the $N_s = 2$ breather as it contains more than one frequency component. A superposition of two solitons can exhibit shape oscillations, but it cannot undergo a transition between single- and double-peak shapes, which requires a superposition of no fewer than three solitons. The breather frequencies are the differences between the chemical potentials, $\mu$, of the constituent fundamental solitons.  Since $\mu \propto (N/N_c)^2$, and the number ratio of the $N_{s} = 3$ breather is 1:3:5 \cite{Satsuma1974}, the ratio of $\mu$ values is 1:9:25, giving frequency ratios of 8:16:24.  Identifying the smallest frequency as $\omega_B$, we have the 3 frequencies: $\omega_B$, $2\omega_B$, and $3\omega_B$, appropriate for $A^{2}=9$.\\
\indent To analyze the 3-soliton breather quantitatively, we fit the integrated 1D-density for each $t_h$ to either a single- or double-Gaussian function depending on whether the central density is a local maximum or minimum, respectively. We extracted the central density $n_{0}(t_{h})$ from the fit, and plot it against $t_h$, as shown by the discrete points in Fig. 3(b). For 3-soliton breathers, $n_{0}(t_{h})$ is fitted to the exact 3-soliton breather solution of the NLSE  for $A^{2}=9$ obtained from the general theory \cite{Satsuma1974, Gordon1983}
\begin{equation}\label{eq:3SolitonDensity}
n_{0}(t_h)=\alpha\left(1+\frac{32[3+5\cos{(\omega_{B}t_h+\phi})]\sin^{2}{\frac{1}{2}(\omega_{B}t_h+\phi)}}{55+18\cos{(\omega_{B}t_h+\phi)}+45\cos{2(\omega_{B}t_h+\phi)}+10\cos{3(\omega_{B}t_h+\phi})}\right),
\end{equation}
with fitting parameters $\omega_{B}$, $\phi$ and $\alpha$. The result is $\omega_B =(2\pi)10.6(1)$ Hz and $\phi=(2\pi)0.11(1)$.  The solid line in Fig. \ref{fig:3SolitonCD}(b) is Eq. (\ref{eq:3SolitonDensity}) using these values. Equation (\ref{eq:3SolitonDensity}) pertains to the specific case of $A^{2}=9$, where the quench produces a pure 3-soliton breather with no radiation. We find that Eq. (\ref{eq:3SolitonDensity}) is a good approximation to the central density of a 3-soliton breather even when $A^{2}$ is close to, but not exactly equal to 9. This result is consistent with exact theory \cite{Satsuma1974} in which a breather composed of three fundamental solitons is created for $6.25<A^{2}<12.25$.\\
\indent In conclusion, we have observed the 2- and 3-soliton breathers in a BEC by quenching the atomic interaction using a zero-crossing of a Feshbach resonance in $^{7}$Li. We have shown that by reducing the axial confinement, the breather frequency approaches the 1D limit, and is well-described by the 1D-NLSE. Like fundamental bright matter-wave solitons, higher-order solitons undergo collapse for a nonlinearity that is too strong. Collapse arises when the soliton is brought close to the 3D boundary, but notably, the collapse threshold for breathers is higher than it is for fundamental solitons with the same total particle number.\\
\indent In the strict 1D limit, breathers are exact solutions of the NLSE. Breathers are superficially similar to time crystals \cite{Sacha2018, Else2020, Yao2020, Autti2018}, although breathers are not a consequence of a \textit{spontaneously} broken symmetry. Breathers are particularly sensitive to beyond-mean-field quantum effects, which, according to Ref. \cite{Yurovsky2017}, lead to formation of a quantum superposition of two fundamental solitons with different relative velocities and numbers of atoms after the quench. Spontaneous dissociation of the breather is predicted to occur after multiple breathing periods \cite{Yurovsky2017, Marchukov2019}. In our experiment, the 2-soliton breather survives for at least two breathing periods. An extension of this work is to measure the breathing duration, which determines if spontaneous dissociation can be observed. Preliminary experiments indicate that noise in the center-of-mass coordinates pose a technical limit to the breather lifetime. Further progress will require better stability of the magnetic field and laser pointing to mitigate center-of-mass fluctuations and drift.\\

\nocite{Pollack2009}
\nocite{Salasnich2006}
\nocite{Lieb1963,McGuire1963,Berezin1964}
\nocite{Yurovsky2008b}
\nocite{Lai1989a}
\nocite{Gammal2001}
\begin{acknowledgments}
This work was supported by the NSF (PHY-1707992, PHY-1912542, and PHY-1607221), joint NSF-BSF (Binational (US-Israel) Science Foundation, Grant No. 2015616), Israel Science Foundation (Grant No. 1286/17)
 and the Welch Foundation (Grant No. C-1133).
\end{acknowledgments}

\newpage
%

\newpage
\newpage
\newpage

\end{document}


\begin{center}
\Large{Supplementary Materials for} \\
\Large{Creation and Characterization of Matter-Wave Breathers} \\
\large{D. Luo, Y. Jin, J. H. V. Nguyen, B. A. Malomed, O. V. Marchukov}\\
\large{V. A. Yurovsky, V. Dunjko, M. Olshanii, R. G. Hulet}\\
\small{Edited Sep 2, 2020}
\end{center}
\subsection*{Error analysis}
The uncertainties in $N/N_{c}$ and in $A^2$ arise from the uncertainties in the measured quantities: $\omega_r$, $N$, and $a$. The radial frequency $\omega_{r}$ is measured by parametric excitation of a trap mode of the BEC at a frequency 2$\omega_r$ which produces observable heating and atom loss. The loss feature is fit to a Lorentzian, giving $\omega_r = (2\pi)297(1)$ Hz. The uncertainty in $N$ is due mainly to a 7\% systematic uncertainty in the imaging laser detuning.  The scattering length is determined from the axial size of the BEC measured as a function of the magnetic field $B$ and compared to a 3D GPE simulation \cite{Pollack2009}. For $B$ between 536 G and 544 G, $a(B)$ is fit to a linear function, $a(B) = \alpha(B-B_0)$, where $\alpha = 0.091(4)$ $a_{0}$/G and $B_{0} = 543.8(2)$ G are the fitted parameters. The uncertainty in $B_0$, arising from the accuracy of the field calibration by RF spectroscopy, results in a systematic uncertainty of 0.02 $a_0$ in $a$, and the uncertainty in $\alpha$ gives an additional fractional uncertainty $\Delta a/a = 4.5$\%, where the former dominates the uncertainty in $a_{i}$, while the latter contribution dominates the uncertainty in $a_f$. The effect of the magnetic dipole interaction (MDI) is included in the calibration of the zero-crossing location, $B_0$ \cite{Pollack2009}. The atomic dipoles are aligned along the $z$-axis which produces an effective attraction for our quasi-1D geometry. In the analysis presented here, however, instead of solving the 3D GPE with MDI, we solve the 1D GPE without MDI. The neglect of MDI produces an effective $B_0$ shifted from the original. We found that the data in Fig. 2(a) are in best agreement with the 1D GPE simulations for $B_{0} = 544.0$ G. We justify using this effective value of $B_{0}$ to evaluate $a(B)$ since it is both consistent with the otherwise neglected MDI shift, and it is within our measurement uncertainty of the unshifted value.

\subsection*{Factorization ansatz for breathers beyond the one-dimensional regime}
Here we use a factorization ansatz to obtain an analytic approximation of the collapse threshold for the 2-soliton breather. 
Consider $N$ atoms with mass $m$ trapped in the harmonic potential with frequency $\omega_{r}$ in the transverse ($xy$) direction. Let $\hbar\omega_{r}$ be the energy unit and $a_{r}=\sqrt{\hbar/m \omega_{r}}$ be the length unit. This system is described by the Schr\"{o}dinger equation
\begin{equation}\label{eq:schrodinger}
N E_{3D}\psi=\hat{H}_{3D}\psi,
\end{equation}
where
\begin{equation}
\hat{H}_{3D} = \sum_{j=1}^{N}\left(-\frac{1}{2}\frac{\partial^{2}}{\partial z_{j}^{2}}+\hat{H}_{\perp}(r_{j})\right)+\frac{4\pi a}{a_{r}}\sum_{j<j'}\delta( \bm{r} _{j}-\bm{r}_{j'}),
\label{eq:H3D}
\end{equation}
$a$ is the scattering length, 
\begin{equation}
\hat{H}_{\perp}=-\frac{1}{2}\left(\frac{\partial^{2}}{\partial r^{2}}+\frac{1}{r}\frac{\partial}{\partial r}\right)+\frac{1}{2}r^{2},
\end{equation} 
and $r_{j}=\sqrt{x_{j}^{2}+y_{j}^{2}}$ is the transverse radius.\\
Following \cite{Salasnich2006}, let us take the wavefunction in the form of a product of many-body axial and transverse wavefunctions
\begin{equation}
\psi=\varphi(\{z\})\prod_{j=1}^{N}\Phi (r_{j}),
\end{equation}
where $\{z\}=\{z_{1},...,z_{N}\}$ is the set of atom axial coordinates and the transverse function is a Hartree product of single-atom functions $\Phi (r_{j})$ of the transverse radius $r_{j}$ (the transverse ground state contais only axially symmetric functions). The  single-atom functions are normalized i.e. 
\begin{equation}
2\pi\int_{0}^{\infty}rdr|\Phi(r)|^{2}=1.
\end{equation}
Projection of the Schr\"{o}dinger equation (\ref{eq:schrodinger}) onto the transverse functions leads to the Lieb-Liniger-McGuire model \cite{Lieb1963,McGuire1963,Berezin1964} for the axial function
\begin{equation}
\left[-\frac{1}{2}\frac{\partial^{2}}{\partial z_{j}^{2}}+\tilde{g}_{1D}\frac{a}{a_{r}}\sum_{j<j'}\delta(z_{j}-z_{j'})\right]\varphi(\{z\})=N E_{\{N\}}\varphi(\{z\}),
\label{eq:schr1D}
\end{equation}
where
\begin{equation}\label{eq:g1D}
\tilde{g}_{1D}=8\pi^2 \int rdr|\Phi(r)|^{4}
\end{equation}
is the effective 1D interaction strength. When $\Phi(r)$ is the ground-state wavefunction of the transverse harmonic potential, we have $\tilde{g}_{1D}=2$ \cite{Yurovsky2008b}, in agreement with the nonlinear coupling constant $g_{1D}$ in Eq. (2) in the main text.
Assuming $a<0$, there exists multistring solutions \cite{McGuire1963}, in addition to the single-string solutions considered in \cite{Salasnich2006}. Due to the translational invariance of the Hamiltonian (\ref{eq:H3D}) and Eq. (\ref{eq:schr1D}) in the $z$-direction, these solutions are also translationally invariant and have homogeneous density. Localized solutions, corresponding to mean-field multi-solitons, can be constructed as a superposition of multistring solutions with different string velocities \cite{Lai1989a}. The mutistring energy tends to the multi-soliton energy in the mean-field limit, and  for the $N_{s}$-soliton breather the energy per atom is given by
\begin{equation}
E_{\{N\}}=-\frac{1}{24}\left(\frac{\tilde{g}_{1D}aN}{a_{r}}\right)^2\epsilon_{\{N\}}.
\end{equation}
Here 
\begin{equation}
\epsilon_{\{N\}} \approx \frac{1}{N^3}\sum_{i}N_{i}^3
\end{equation}
and the numbers of atoms in the constituent solitons are $\{N\}=\{N_{1},N_{2},...,N_{N_{s}}\}$ with 
$\sum_{i}N_{i}=N$.\\
The transverse single-atom functions  can be evaluated using the variational principle for the total energy 
$\langle\psi|\hat{H}_{3D}|\psi\rangle=N\left(E_{\{N\}}+\langle\Phi|\hat{H}_{\perp}|\Phi\rangle\right)$. Unlike the Gaussian variational function, used in \cite{Salasnich2006}, here the variation over $\delta\Phi^*$ leads to the radial GPE
\begin{equation}\label{eq:radialGPE}
\left[\hat{H_{\perp}}-\frac{16}{3}\pi^3\left(\frac{a}{a_{r}}\right)^{2}N^{2}\epsilon_{\{N\}}\int r'dr'|\Phi(r')|^{4}|\Phi(r)|^{2}\right]\Phi(r)=E_{r}\Phi(r).
\end{equation}
It depends only on the universal parameter --- the scaled atom number
\begin{equation}\label{eq:tilN}
\tilde{N}=\frac{a}{a_{r}}\sqrt{\epsilon_{\{N\}}}N
\end{equation}
and was solved numerically.  The solution diverges showing collapse at $\tilde{N}\ge 0.717$. Therefore, a collapse occurs at $N>N_{c}/\sqrt{\epsilon_{\{N\}}}$, where $N_{c}=0.717a_{r}/a$ is the critical number of atoms for the single string, corresponding to the fundamental soliton. The factor of 0.717 is closer to the value of 0.676, obtained in \cite{Gammal2001} by a numerical solution of 3D GPE, than the value of 0.76 in \cite{Salasnich2006} with the Gaussian transverse function.\\
The critical atom number depends on the axial state since the effective 2D interaction strength in (\ref{eq:radialGPE}) is proportional to the binding energy of the multi-soliton state. Then the collapse threshold increases with the number of solitons. For $N_{s}$-breather containing solitons with masses $N_{i}=(2i-1)N/N_{s}^{2}$ $(1\leq i\leq N_{s})$ we have 
$\epsilon_{\{N\}}\approx(2N_{s}^2-1)/N_{s}^4$ and, therefore, collapse is predicted at 
$N/N_{c}>N_{s}^2/\sqrt{2N_{s}^2-1}$.  For the $N_{s}=2$ breather, the model gives the approximate estimate of $N/N_{c}=1.5$.

\newpage
%